\documentclass[twocolumn,showpacs,aps,pre]{revtex4}
\usepackage{amssymb}
\usepackage{graphicx}

\begin{document}

[Phys. Rev. E {\bf 73}, 056124 (2006)]

\title{From Scale-free to Erdos-R\'enyi Networks}

\author{Jes\'us G\'omez-Garde\~nes$^{1,2}$ and Yamir Moreno$^{1,3}$}

\affiliation{
$^{1}$Instituto de Biocomputacion y F\'{\i}sica de los Sistemas
  Complejos (BIFI), Universidad de Zaragoza, Zaragoza E-50009, Spain.\\
$^{2}$Departamento de F\'{\i}sica de la Materia Condensada and \\
Instituto de Ciencia de Materiales de Arag\'on (ICMA),
Universidad de Zaragoza - CSIC, Zaragoza E-50009, Spain.\\
$^{3}$Departamento de F\'{\i}sica Te\'orica, Universidad de Zaragoza,
Zaragoza E-50009, Spain.}

\begin{abstract}
We analyze a model that interpolates between scale-free and
Erdos-R\'enyi networks. The model introduced generates a
one-parameter family of networks and allows to analyze the role of
structural heterogeneity. Analytical calculations are compared with
extensive numerical simulations in order to describe the
transition between these two important classes of networks. Finally,
an application of the proposed model to the study of the percolation
transition is presented.
\end{abstract}

\maketitle

\section{Introduction}

In the last several years, graph theory has experienced a burst of activity as
many real systems can be represented and modeled as networks
\cite{bornholdtbook,newmanreview,ourreport}. A network is made up of vertices
representing the interacting elements of the system and of edges that stand
for the interactions among them. Network modeling comprises the analysis and
characterization of the structure of networks as well as their modeling in
terms of generic models aimed at reproducing the features found in real
systems \cite{newmanreview,ourreport,strogatzreview}. The second important
branch has to do with dynamics on networks. This has lately attracted the
attention of many scientists as it is ultimately related with the functioning
of the system that is being modeled \cite{ourreport,strogatzreview}.

In this paper, we deal with the first of these areas of research. The seminal
paper by Barab\'asi and Albert \cite{bascience99,barabasi-1999}, showed that
many real world networks can not be described by graphs where the connectivity
distribution (i.e., the probability that a given node has a given number of
links) follows a Poisson-like distribution. Indeed, Barab\'asi and Albert
showed that most real networks are heterogeneous in the sense that the
probability that a node is connected to $k$ other nodes follows a power-law
distribution $P(k)\sim k^{-\gamma}$, where $\gamma$ usually lies between 2 and
3. These networks were termed scale-free networks
\cite{bornholdtbook,newmanreview,ourreport,strogatzreview,vespignanibook}.

Soon afterwards, many studies have dealt with the analysis and
characterization of models that generate scale-free networks, along with other
global and local topological properties found in real networked systems
\cite{newmanreview,ourreport,vespignanibook,mendesreview,fang}. In particular,
models based on the mechanism of preferential attachment (PA), no matter if
the network is growing or not, have been extensively studied in the last
years. There are some models in which the PA rule is limited to a neighborhood
due to geographic constraints \cite{amaral} or lack of global knowledge
\cite{gardenes-2004}, or where its linear character is investigated
\cite{redner}. While today we have recognized that preferential attachment is
not a necessary condition for the formation of scale-free networks
\cite{cald}, it seems to be clear that it is an important mechanism. Indeed,
most of the existing models intrinsically incorporate a preferential
attachment like rule. On the other hand, uniform random linking of nodes on
growing networks gives rise to networks where the degree distribution decays
exponentially fast with the degree $k$, thus producing homogeneous networks
with a well defined (and meaningful) average value for $k$
\cite{ws98,barratweigt}.

The combination of the two rules, i.e, uniform and preferential linking, have
been also analyzed in several models for interpolating between scale-free and
exponential networks \cite{note1}. For instance, Liu {\em et al} \cite{liu}
have studied a model in which the probability of establishing new links goes
as a linear combination of both in such a way that a new link is established
between a node $i$ and a new one proportionally to $(1-p)k_i+p$, where $p$
weights the contribution of the two mechanisms. However, in previous models of
this sort, there is an assumption that does not apply always. It has to do
with the fact that the network always grows around a single component of
connected nodes and uniform or preferential links from the emerging nodes are
always made with elements belonging to this unique cluster. This single
component grows linearly in time until it reaches the size of the
network. Since there are no clusters of nodes other than the giant component,
the models can not account for phenomena such as the coalescence of small
networks into a larger one, nor for situations in which more than one node is
added to a preexisting structure at each time step, features that may be
relevant in social, economic and other networked systems.

In this paper, we analyze a model that interpolates between Erdos-R\'enyi (ER)
and scale-free (SF) networks as far as the degree distribution is concerned
through a tunable parameter. By construction, new links are not always
established with nodes previouly incorporated to the network. We explore
analytically and numerically the time behavior of nodes attachment as
well as of the degree evolution. We find that, depending on the interplay
between uniform and preferential linking, the transition from an ER like
network to an SF one is smooth or more abrupt. We finally discuss other
topological properties and perform a numerical percolation study that
highlights the differences in their structure. The model presented here is
useful as it provides a unique recipe to go progressively from homogeneous to
heterogeneous topologies as well as for exploring the interplay between them.

\section{The model}

\begin{figure}[tbp]
\begin{center}
\begin{tabular}{c}
\includegraphics[angle=0,width=.25\textwidth]{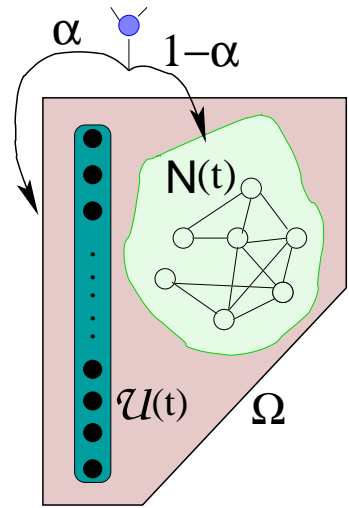}
\end{tabular}
\end{center}
\caption{(Color online). Schematic representation of the general procedure for
generating the networks. With probability $\alpha$ one of the $m$ links can be
made with any of the nodes (and with the same uniform probability) that will
take part in the final network. On the other hand, with probability
$(1-\alpha)$ the link will be made only with those nodes that form the
connected set at that time because the node will choose a preferential linking
strategy.}
\label{fig:1}
\end{figure}

The model introduced in this work generates a one-parameter family of
complex networks. This parameter, $\alpha\in [0,1]$, measures the
degree of heterogeneity of the final networks. Let us assume the final
size of the network to be $\Omega$. The network is generated in the
following way:
\begin{itemize}
\item[{\it (i)}] Start from a fully connected network of $m_{0}$ nodes and 
a set ${\mathcal U}(0)$ of $(\Omega-m_{0})$ unconnected nodes.
\item[{\it (ii)}] At each time step choose a new node $j$ from ${\cal U}(0)$.
\item[{\it (iii)}] This node makes a link in two ways:
\begin{itemize}
\item[(a)] With probability $\alpha$ it links to any other node $i$ of
the whole set of $\Omega-1$ nodes with uniform probability
\begin{equation}
\Pi^{\text{uniform}}_{i}=(\Omega-1)^{-1}\;.
\label{eq:1}
\end{equation}
\item[(b)] With probability $(1-\alpha)$ establish a link following 
a preferential attachment strategy, that is, the probability for any other 
node $i$ to attach to node $j$ is a function of its connectivity as,   
\begin{equation}
\Pi^{\text{PA}}_{i}={\bf F}(k_{i})\;,
\label{eq:2}
\end{equation}
where different choices for the functional form of ${\bf F}(x)$ are
analyzed below.
\end{itemize}
\item[{\it (iv)}] Repeat $m$ times step {\it (iii)} for the same node $j$.
\item[{\it (v)}] Repeat ${\cal U}(0)=(\Omega-m_{0})$ times steps {\it (ii)} 
to {\it (iv)}.
\end{itemize}
A schematic plot of the linking procedure at step {\it (iii)} is shown in
Fig. \ref{fig:1}. The above rules allow for the coexistence of two classes of
nodes. On one hand, there are nodes with at least one link. This set will be
referred to henceforth as the connected set $N(t)$ \cite{note2}. On the other
hand, there is another set ${\cal U}(t)$ of isolated nodes such that its size
is $\Omega-N(t)$. At variance with other models in which there are only nodes
with connectivity different from zero and thus the connected component grows
linearly with time, the above rules allows the addition of more than one node
to the set $N(t)$ as a result of random linking. Therefore, we expect the time
dependency of $N(t)$ to be highly non trivial.

\section{Network growth and degree evolution}

In order to describe the evolution of the nodes degree, one has
to consider the functional form of ${\bf F}(x)$ for the preferential
attachment probability (\ref{eq:2}). However, we can take into account
some previous considerations that do not depend on the particular form of
${\bf F}(x)$.

First of all, it is useful to consider two kind of links in order to analyze
the model. Namely, the ones that arise from a uniform random choice,
$k^{\text{u}}$, and the remaining, $k^{\text{pa}}$, corresponding to the
implementation of the preferential attachment rule.  The dynamics of
$k^{\text{u}}$ is completely independent of the dynamics of the PA links,
$k^{\text{pa}}$, but the opposite is not necessarily true. From this, it
follows that the probability that one node has $k^{\text{u}}$ uniform links,
$P^{\text{u}}(k^{\text{u}})$, is a Poisson distribution with $\langle
k^{\text{u}} \rangle=2\alpha m$.
\begin{equation}
P^{\text{u}}(k^{\text{u}})=\frac{(2\alpha m)^{k^{\text{u}}}{\mbox e}^{-2\alpha
m}}{k^{\text{u}}!}
\label{eq:5}
\end{equation}
As a consequence, we will concentrate on analyzing the growth dynamics
of the PA links for the studied models.

\begin{figure}[tbp]
\begin{center}
\begin{tabular}{c}
\includegraphics[angle=0,width=.50\textwidth]{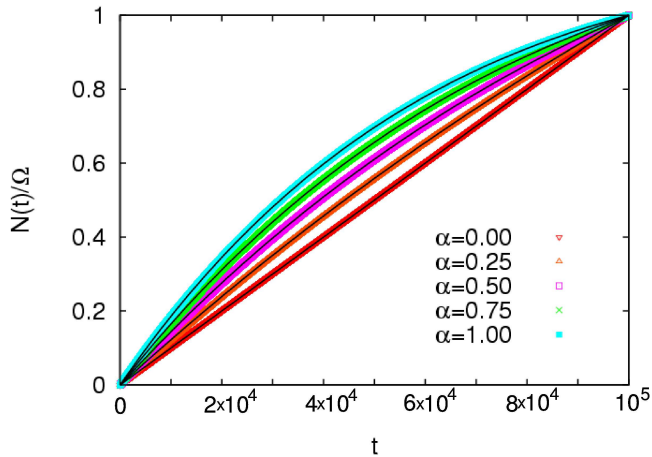}
\end{tabular}
\end{center}
\caption{(Color online). Size of the connected set $N(t)$ as a function of
time.  Solid lines correspond to the analytical results (eq. (\ref{eq:4})) and
points are the Monte Carlo results of network construction (employing model A
(sec. \ref{sec:ModelA})). The comparison is made for $\Omega=10^5$ and several
values of $\alpha$. The parameters of the model are set to $A=m=m_{0}=1$.}
\label{fig:2}
\end{figure}

It is particularly interesting to study at this point how uniform random
linking affects the evolution of the connected set since this is
completely independent on the specific PA rule considered. This feature
represents one of the main differences between the studied model and other
previous mechanisms used to generate growing networks
\cite{newmanreview,ourreport,mendesreview}. That is, in our model nodes are
not incorporated to the connected set at a constant rate (like {\em
e.g.} in the standard Barab\'asi-Albert model) due to the possibility of
adding new nodes from ${\cal U}(t)$ by applying uniform linking at time $t$ and
therefore the set ${\mathcal U}(t)\neq{\mathcal U}(0)-t$.  We can easily
derive the evolution of the connected set size, $N(t)=\Omega-{\mathcal
U}(t)$, for any value of the parameter $\alpha$.  For this, we consider the
growth of the connected set at each time step, {\em i.e.} when a new
node of ${\mathcal U}(0)$ throws its $m$ links
\begin{equation}
N(t+1)=N(t)+\frac{\Omega-N(t)}{\Omega-(t+m_{0})}
       +\alpha m \left (1-\frac{N(t)}{\Omega}\right )\;.
\label{eq:3}
\end{equation}
In the above equation the second term on the right accounts for the
probability that the new node (which is throwing its $m$ links) of ${\mathcal
U}(0)$ does not belong already to the connected set at time $t$ (due to the
possible uniform links obtained from previous nodes of ${\mathcal U}(0)$
already connected to the connected set $N(t)$). Besides, the third term on the
right describes the probability that any uniform link thrown by the node is
directed to a node belonging to ${\cal U}(t)$. These two terms account for the
growth rate of the connected set. We can consider that both time and $N(t)$
are continuous variables and make the time step small enough in order to
obtain the corresponding ODE associated to eq. (\ref{eq:3}), whose solution is
given by
\begin{equation}
N(t)=\Omega+(t+m_0 -\Omega){\mbox e}^{-\alpha m t / \Omega}\;.
\label{eq:4}
\end{equation}   
The agreement between this calculation and Monte Carlo simulations is
shown in Fig. \ref{fig:2} for different values of $\alpha$ and a
preferential attachment as described in what follows (model A)
\ref{sec:ModelA}. It is worth noting the highly nonlinear behavior of
$N(t)$, at variance with models in which its size changes at a
constant rate.

We formulate below two different ways to implement the preferential
attachment rule, which give rise to different behaviors. In both
models we will consider that the PA probability of a node $j$ depends
only on the PA links of the node, ${k}_j^{\text{pa}}$. This new separation
between PA links and uniform ones introduces a higher differentiation
between the two simultaneous kinds of link dynamics implemented here
allowing us to manipulate (as shown below) the degree of correlation
between them. The two models interpolate between scale-free and
Erdos-R\'enyi topologies but the structural transition is quite
different (as we will show in section \ref{sec:Statistics}).

\subsection{MODEL A}
\label{sec:ModelA}

\begin{figure}[tbp]
\begin{center}
\begin{tabular}{c}
\includegraphics[angle=0,width=.50\textwidth]{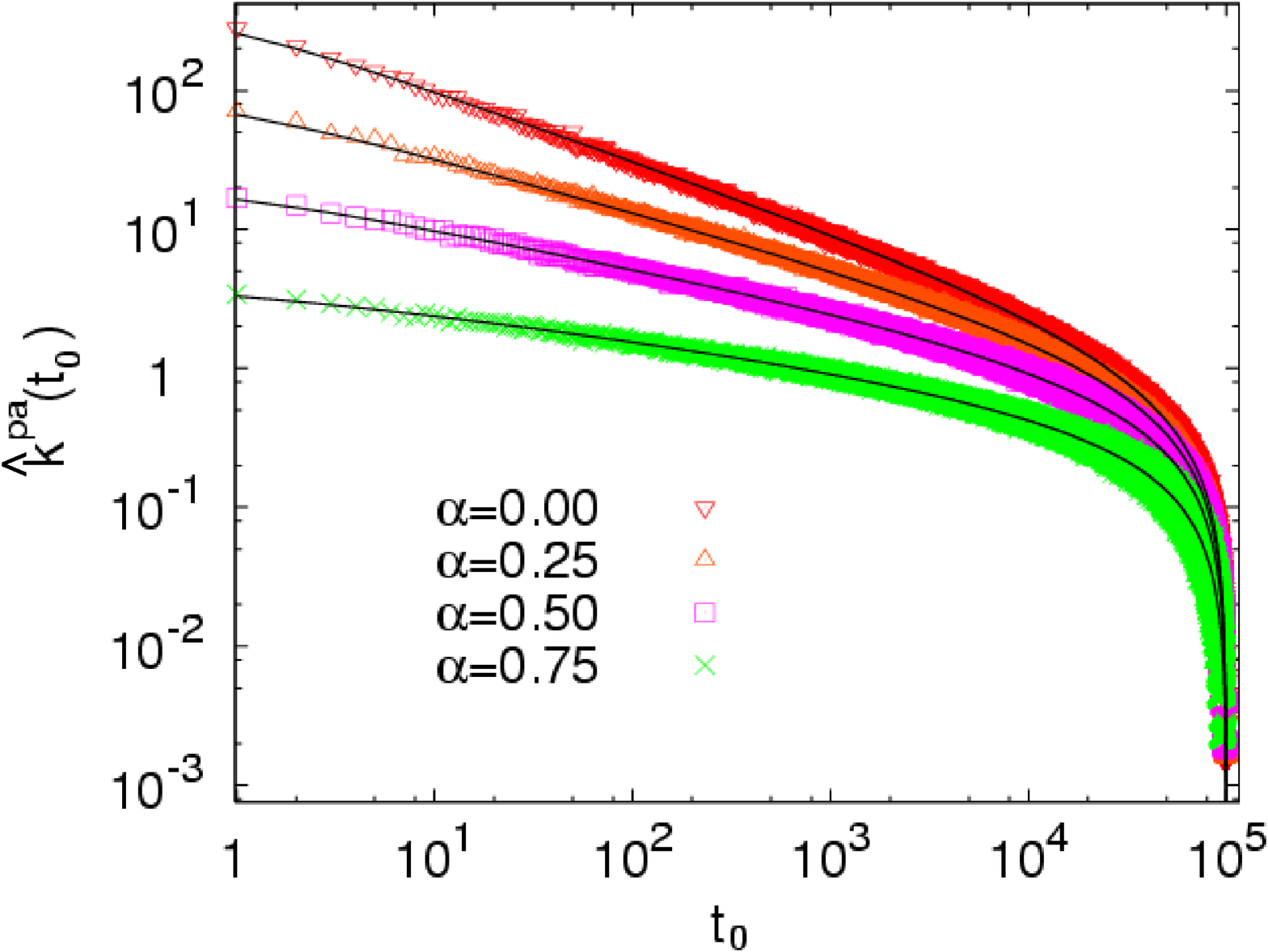}
\end{tabular}
\end{center}
\caption{(Color online). {\bf Model A}. Monte Carlo simulation (points) versus
mean field (lines) results for $\hat{k}^{\text{pa}}(t=\Omega)$ as a function
of the birth time $t_{0}$ for different values of $\alpha$. The parameters of
the model were $\Omega=10^5$ and $A=m=m_{0}=1$. The statistics of the Monte
Carlo simulations were performed using $10^4$ networks for each value of
$\alpha$.}
\label{fig:3}
\end{figure}

In this first model we shall study a preferential attachment rule strongly 
correlated with the simultaneous uniform random linking. 
First, we consider that the PA probability of a node $i$ is linear with the
{\em incoming} PA degree of the node, $\hat{k}^{\text{pa}}_{i}$, that is,
those links received by $i$ when other node launches (in average)
$(1-\alpha)m$ links following the PA rule.  This particularity of the PA rule
was already considered by Dorogovtsev {\em et al} \cite{dorogovtsev-2000}.
Besides, we consider that when a node is introduced in the connected
component (because either it is chosen at random by any node or it is
launching its $m$ outgoing links over the rest of nodes) it has an
initial attractiveness (or fitness) $A$. In other words, each node has
an associated parameter $A_{i}$ that is zero if the node $i$ is not in
the connected set and is $A_{i}=A$ if $i$ is linked to other
nodes ({\em i.e.}, it belongs to $N(t)$). We further consider that the
attractiveness $A_{i}$ enters linearly in the preferential linking
probability of node $i$. With these two ingredients, the expression
for $\Pi^{\text{PA}}_{j}$ is given by
\begin{equation}
\Pi^{\text{PA}}_{i}=\frac{\hat{k}^{\text{pa}}_i+A_i}{\sum_{j\in\Omega}(\hat{k}^{\text{pa}}_j+A_j)}\;,
\label{eq:8}
\end{equation}
The introduction of the fitness $A$ correlates the PA rule with the uniform
linking in the sense that the more links are {\bf established uniformly} (the
higher $\alpha$), the more new nodes with $\hat{k}^{\text{pa}}_{i}=0$ are
incorporated to the connected set from ${\cal U}(t)$ and hence (by the
presence of $A$ in the PA probability) the more candidates to obtain PA links
are available. This can be observed from the evolution of the connected
set $N(t)$, when $\alpha$ is high there are a lot of nodes added into
$N(t)$ at the early stage of the network construction so that the potential
growth of the PA degree of the former members of the connected set is
strongly weakened. In order to confirm these heuristic considerations we
derive the mean field evolution for the incoming PA degree of a node $i$,
$\hat{k}^{\text{pa}}_i$
\begin{equation}
\frac{{\mbox d}\hat{k}^{\text{pa}}_i}{{\mbox d}t}=
(1-\alpha)m\frac{\hat{k}^{\text{pa}}_i+A}{(1-\alpha)mt+A\;N(t)}\;,
\label{eq:MFMA}
\end{equation}
(with the initial condition $\hat{k}^{\text{pa}}_i(t_{0}^i)=0$). Obviously, in
the limit $\alpha=0$ we recover the mean field equation for the Generalized
Dorogovtsev model \cite{dorogovtsev-2000} (which, when $A=m$, describes the
Barab\'asi-Albert model). For $\alpha\neq0$ the influence of the uniform
random linking is evident from the presence of $N(t)$. The number of nodes
that start to have the above dynamics at some time $t_0$ is
${\mbox{d}}N(t)/\mbox{d}t$ evaluated at time $t=t_0$ which for $\alpha\neq0$
is not constant as we have seen in the previous calculation of $N(t)$. The
solution of (\ref{eq:MFMA}) is then given by
\begin{equation}
\frac{\hat{k}^{\text{pa}}_i(t=\Omega)}{A}=
-1+{\mbox{exp}}\left[(1-\alpha)m\int_{t_{0}^i}^{\Omega}
\frac{{\mbox d}t}{(1-\alpha)mt+A\;N(t)}\right]\;.
\label{eq:MFMAsol}
\end{equation}

We have solved numerically eq. (\ref{eq:MFMAsol}) in order to obtain
$\hat{k}^{\text{pa}}_{i}(t=\Omega)$ (or $k^{\text{pa}}_{i}(t=\Omega)=
\hat{k}^{\text{pa}}_{i}(t=\Omega)+\alpha m$) as a function of $t^{i}_0$. This
function, along with the number of nodes which are incorporated to the
connected set at time $t_{0}^i=t_{0}$, gives the degree
distribution of the PA links. We have compared the results given by
eq. (\ref{eq:MFMAsol}) for different values of $\alpha$ with the
corresponding ones obtained by performing Monte Carlo simulations of
the model (averaging over $10^4$ networks for each value of
$\alpha$). The results, plotted in Fig. \ref{fig:3}, show a very good
agreement for the mean field model and the numerical network
construction.  As expected, the sooner a node is incorporated to the
connected set the higher its final PA degree. However, as
discussed above, one can observe that this gain of the oldest nodes
becomes less important when the value of $\alpha$ grows due to the
combination of two effects: {\it(i)} the application of the PA rule
becomes less frequent and {\it(ii)} the fast growth of the connected
set tends to make more homogeneous the PA probability of the
nodes.

\subsection{MODEL B}

In the second  proposal the two different linking processes are 
completely independent. For this, we consider that $\Pi_{i}^{\text{PA}}$ 
is a linear function of the (incoming and outgoing) links 
that appear as a product of the application of the PA rule. 
Then, $k_{i}^{\text{pa}}$ will be zero until it launches its 
$\alpha m$ PA links over the rest of the nodes, {\em i.e.} 
regardless of $k_{i}^{\text{u}}$. Then, the mean field equation for the 
evolution of $k_{i}^{\text{pa}}$ is given by
\begin{equation}
\frac{{\mbox d}k_{i}^{\text{pa}}}{{\mbox d}t}=
(1-\alpha)m\frac{k^{\text{pa}}_{i}}{2(1-\alpha)mt+m_0}\;,
\end{equation}  
with the initial condition $k^{\text{pa}}_{i}(t_{0}^{i})=(1-\alpha)m$ and
$t_{0}^{i}$ being the time when node $i$ launches its $m$
links. Solving the above equation yields
\begin{equation}
k_{i}^{\text{pa}}(t)=(1-\alpha)m\left[\frac{t}{t_{0}^i}\right]^{1/2}\;.
\end{equation}
Because the nodes launch their links at a constant rate (one node per
time step), it is easy to obtain the degree distribution $P(k^{\text{pa}})$
\begin{equation}
P(k^{\text{pa}})=2(1-\alpha)^2m^2(k^{\text{pa}})^{-3}\;, 
\end{equation}
which is simply a power law distribution with a Barab\'asi-Albert
exponent regardless of the value of $\alpha$.  On the other hand, the
relative weight of the power law with respect to the Poisson
distribution in the total degree distribution $P(k)$ will be obviously
affected by $\alpha$ (as the prefactor in the above equation
suggests).

\section{Network properties}

\begin{figure*}[tbp]
\begin{center}
\begin{tabular}{c}
\includegraphics[angle=0,width=.90\textwidth]{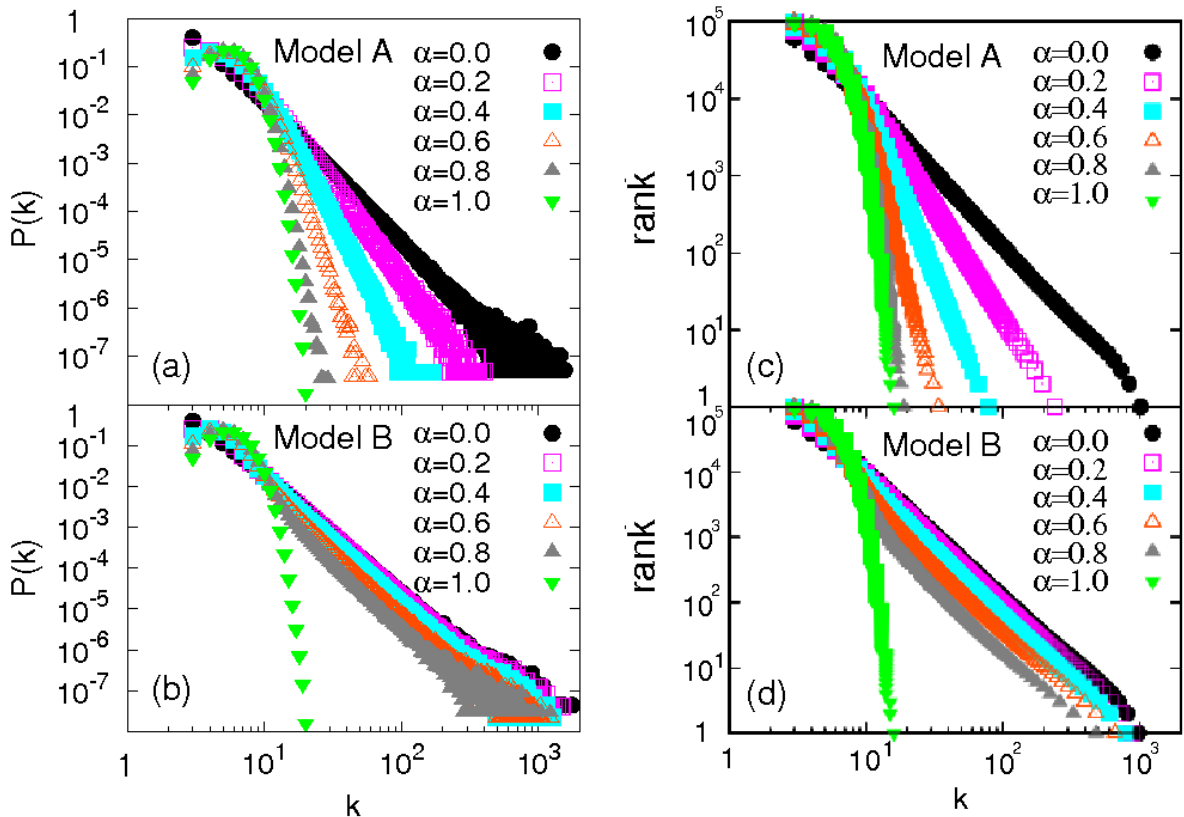}
\end{tabular}
\end{center}
\caption{(Color online). Monte Carlo results for the degree distribution
$P(k)$ and rank-degree relation for several values of $\alpha$. ({\bf a}) and
({\bf c}) show the results for model A revealing a progressive increase of the
tails decaying rate when $\alpha\rightarrow 1$. The results for model B (({\bf
b}) and ({\bf d})) show how the decaying rate is not affected by $\alpha$. The
networks were generated with the following parameters $\Omega=10^5$ and
$m=m_{0}=3$ ($A=3$ for model A).}
\label{fig:4}
\end{figure*}

\begin{figure}[tbp]
\begin{center}
\begin{tabular}{c}
\includegraphics[angle=0,width=.45\textwidth]{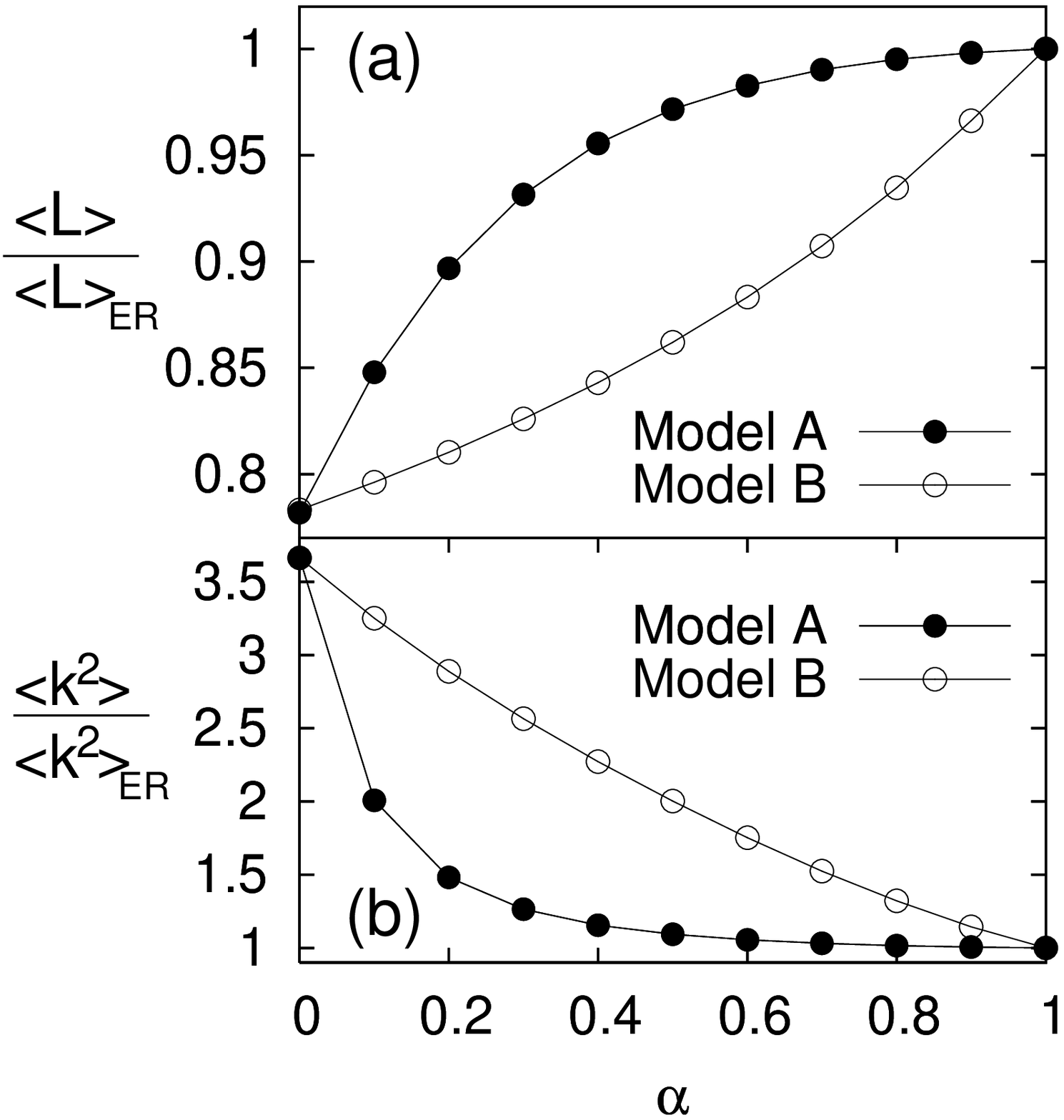}
\end{tabular}
\end{center}
\caption{Average path length {\bf (a)} and second moment of 
the degree distribution {\bf (b)} as a function of $\alpha$. 
Both quantities are represented normalized by their respective 
values in the ER limit. The results clearly manifest the two 
different transitions of the models regarding the heterogeneity 
evolution along the interpolating path. The averaged networks 
had the following parameters $\Omega=10^4$ and $m=m_{0}=3$ 
($A=3$ for model A).}
\label{fig:5}
\end{figure}

In this section we discuss the transition from SF to ER networks in
terms of the global topological features of the networks. We have
performed Monte Carlo simulations of the two models and compared how
the relevant topological measures evolve as a function of $\alpha$. We
are interested in obtaining how the different correlations between the
uniform and PA linking rules affect several structural measures.  To do
this, we have studied the behavior of three magnitudes that behave
very different in the two known limiting cases (SF and ER networks),
namely: the degree distribution $P(k)$, the average shortest path
length $\langle L\rangle$ and the second moment of the degree
distribution $\langle k^2\rangle$.

{\em Degree distribution} - The degree distribution evolution is clearly
different for the two models. In Fig. \ref{fig:4} we have plotted the degree
distribution and the rank-degree relation for both models. The rank-degree
relation provides a useful tool for analysing the degree heterogeneity of the
networks \cite{keller} and thus it is helpful when looking at the transition
between ER and SF networks.  As can be observed from Figs. \ref{fig:4}(a) and
\ref{fig:4}(c) the correlated model A shows a smooth transition from the power
law ($\alpha=0$) to the Poisson distribution ($\alpha=1$). The decay of the
tails ($k>>1$) of the degree distribution and the rank-degree relation becomes
progressively faster as $\alpha$ grows revealing the decrease of the exponent
of $P^{\text{pa}}(k^{\text{pa}})$ as expected from the results obtained by the
analytical insights developed for model A.  For model B the transition is
completely different as it is shown in Figs. \ref{fig:4}(b) and
\ref{fig:4}(d). In both representations the decaying rate of the tails is
independent of $\alpha$ and the transition to the Poisson distribution is much
more apparent for low values of $k$. In this sense one can conclude that
highly connected nodes persist along the transition of model B while for model
A the heterogeneity is progressively lost.

{\em Average shortest path length} - The different evolution of the
degree distributions observed above suggests to look at how the
average shortest path length behaves along the two paths of
interpolation.  It is well known that the existence of high degree
nodes makes the network more compact due to the possibility of finding
shortcuts between nodes going through the hubs. Hence, the persistence
of highly connected nodes determines the small diameter of the
scale-free network. The results obtained are shown in
Fig. \ref{fig:5}(a).  As expected, the average shortest path length as a
function of $\alpha$ grows slower for model B because the probability
of finding hubs is higher than for the networks generated using model
A for the same value of $\alpha$.

{\em Second moment of $P(k)$} - In order to obtain a quantitative
measure of the evolution of the degree heterogeneity for the two
models it is convenient to measure the second moment of the degree
distribution, $\langle k^2\rangle$. This magnitude diverges (in the
thermodynamic limit $\Omega\rightarrow\infty$) for scale-free networks
with exponents between 2 and 3. So, we expect a decrease of the
heterogeneity on the path to ER graphs. As can be observed from
Fig. \ref{fig:5}(b), model A shows a faster decrease of $\langle k^2\rangle$
as expected from the study of the degree distribution while for model
B the transition is much smoother revealing again the persistence of
highly connected nodes along the path to the ER limit.

As for other properties like the clustering coefficient and
degree-degree correlations we have checked that they remain unchanged
irrespective of the value $\alpha$ and wheter model A or B is
implemented.

\section{Percolation dynamics}
\label{sec:Statistics}

\begin{figure}[tbp]
\begin{center}
\begin{tabular}{c}
\includegraphics[angle=0,width=.45\textwidth]{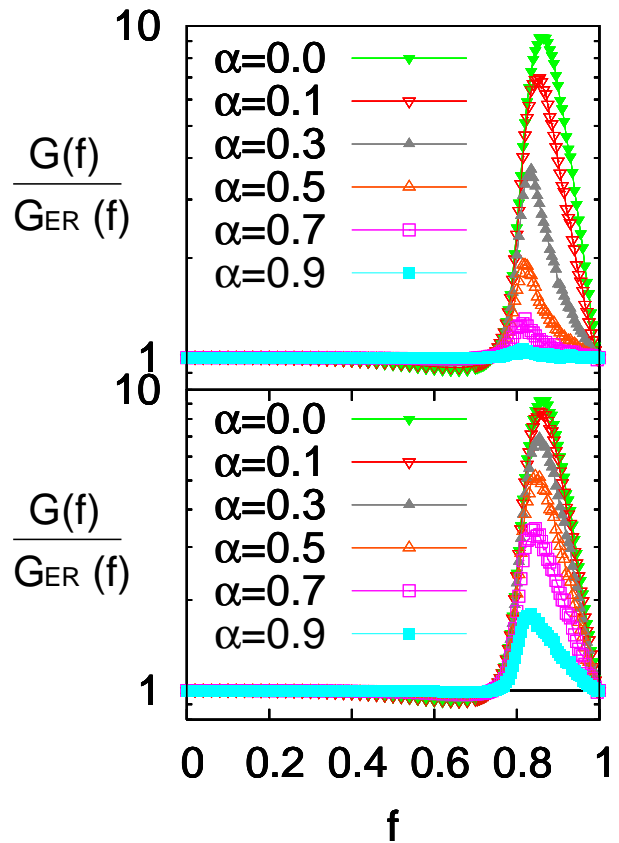}
\end{tabular}
\end{center}
\caption{(Color online). Ratio between the size of the giant connected
  component in networks generated using models A (upper) and B (bottom) and ER
  networks when a fraction $f$ of nodes is randomly deleted for different
  values of $\alpha$. Note that although the transition points for models A
  and B are locate close to each other, the dependence with $\alpha$ is quite
  different. Network parameters are those of Fig \ref{fig:5}.}
\label{fig6}
\end{figure}

One of the most important differences between ER graphs and SF
networks is given by the radically different behaviors of dynamical
processes that take place on top of them
\cite{cnsw00,ceah00,ceah01,pv00,pv01,mpv02}. For instance, epidemic
spreading processes show a natural threshold below which the epidemic
can not spread for ER graphs, while this threshold is absent in the
thermodynamic limit in SF networks with a diverging second moment
\cite{pv00,pv01,mpv02}. This kind of behavior is precisely what makes
scale-free networks so special.

We have implemented a percolation process on top of the networks
generated by models A and B. It is aimed at simulating the random
failures of a fraction $f$ of nodes \cite{cnsw00,ceah00}. By computing
the size of the giant connected component, one can characterize the
percolation transition. Here, however, we are not interested in the
transition {\em per se}, but on the influence of the topological
features unraveled in the previous section on the size of the giant
component of the network. As expected, the behavior of the two models
is different when the limit of ER graphs is approached for the same
values of $\alpha$.

As usual, we have analyzed the evolution of the size of the giant component of
the network after a fraction $f$ of the nodes (and hence their links) are
removed from it. In this sense, it is relevant to study the relation
$G(f)/G_{\text{ER}}(f)$, where $G(f)$ ($G_{\text{ER}}(f)$) is the size of the
giant component of the network for $\alpha\neq1$ ($\alpha=1$), since this will
clearly unravel the two different approaches from the scale-free limit to the
Erdos-R\'enyi network. As can be observed from Fig. \ref{fig6} the differences
between the SF and ER networks are relevant when $f$ is high due to the
different critical behaviors near the transition point
\cite{cnsw00,ceah00}. More interesting for our concerns is the evolution of
the magnitude $G(f)/G_{\text{ER}}(f)$ as a function of $\alpha$. For high
values of $f$ we can clearly appreaciate the fast approach from SF to ER
supplied by model A (upper plot of Fig. \ref{fig6}). In fact, the size of the
giant component is very similar to that corresponding to the ER graph for
($\alpha>0.7$). In contrast, the transition exhibited for the same values of
$f$ when model B is implemented reveals again a slower transition between SF
and ER networks (Fig. \ref{fig6} down). In particular one can observe that the
values of $G(f)/G_{\text{ER}}(f)$ for $\alpha=0.9$ are similar to those for
$\alpha=0.5$ in model A.

The two different behaviors observed can be explained in terms of the
topological measures showed in the previous section, in particular with the
behavior of the second moment of the degree distribution $\langle k^2\rangle$
which is known to play a key role for understanding the different transitions
in SF and ER networks. It is known that the critical value of $f$ relates to
the moments of the distribution as $1-f_c=1/\kappa_0$, where
$\kappa_0=(\langle k^2 \rangle/\langle k \rangle)-1$. From Fig.\
\ref{fig:5}(b), one can see that for model A, the second moment of the degree
distribution approaches the value obtained in ER limit for intermediate values
of $\alpha$. This is no anymore the case for model B, in which the ratio
$\langle k^2 \rangle/\langle k^2 \rangle_{\text{ER}}$ approaches $1$ only for
values close to $\alpha=1$. Given that the average connectivity $\langle k
\rangle$ is the same for all values of $\alpha$ in both models, we should
expect that the behavior of a percolation process near the critical point
would be nearly the same to that of ER networks when $\langle k^2
\rangle/\langle k^2 \rangle_{\text{ER}}\rightarrow 1$. This is the situation
for a broader range of $\alpha$ values in model A as suggested by Fig.\
\ref{fig:5}(b). In other words, the different behaviors for the two networks
families observed in Fig.\ \ref{fig6} can be explained in terms of the speed at
which $\langle k^2 \rangle$ convergences to its corresponding value in the ER
limit.

The implemented percolation dynamics serves us to illustrate how the
interpolating model presented in this work can be a useful tool for
discovering what topological features are fundamental to explain the
different behaviors observed when more complex dynamics are
studied. Moreover, real networks are not purely scale-free nor
completely ER, so that their behavior in front of percolation like
processes lies in between these extreme cases as clearly shown in
Fig.\ \ref{fig6}.

\section{Conclusions}

In this paper, we have analyzed a model that interpolates between
Erdos-R\'enyi and scale-free networks. The combination of uniform and
preferential linking allows us to explore the whole path between the two
limiting cases. An important feature of our model is that the size of the
connected set does not grow linearly with the number of nodes attached to the
network. This is a result of the novel ingredient of the model, that through
the uniform random linking allows every node to take part on the network
irrespective of their connectivities (even if they are not connected at
all). We have analyzed two different variants for the interpolation between
the Erdos-R\'enyi and scale-free limits. On one model the transition is
smooth, while for the other it becomes sharper. The analytical insights
together with numerical simulations supported that the differences in the
versions analyzed are rooted at the interplay between uniform and preferential
attachment. Finally, simulations of a percolation process have illustrated the
differences in both formulations and their associated transitions (which are
easily explained when looking at the topological properties of the networks).

As for future works, the present model provides a useful tool to study
the influence of the degree of heterogeneity in dynamical processes of
different kinds just as the Watts-Strogatz model have proved to do so
in the transition from regular to random structures. In particular,
there exist open questions in phenomena such as the synchronization of
coupled oscillators \cite{grinstein} where this kind of model could be
particularly relevant to explore the system's behavior in the region
where homogeneous and heterogeneous architectures coexist.

\begin{acknowledgments}
  J.\ G.-G.\ acknowledge financial support of the MEC through a FPU grant. Y.\
  M.\ is supported by MEC through the Ram\'{o}n y Cajal Program.  This work
  has been partially supported by the Spanish DGICYT Projects
  FIS2004-05073-C04-01 and FIS2005-00337 .
\end{acknowledgments}


\begin{thebibliography}{99}
\bibitem{note1} Note that all the networks studied here are
  random. The only difference is that they are obtained through two
  distinct attachment process, namely, uniform and nonuniform random
  linking.
\bibitem{note2} We recall on the possibility that the connected set is
  temporarily composed by more than one connected component. This is the case
  for the initial stage of the network growth when $\alpha\rightarrow
  1$. However, when $\Omega$ is high enough the final network is composed by a
  unique connected component.
\end{thebibliography}
\end{document}